\documentclass[useAMS]{mn2e}
\usepackage{psfig}

\usepackage{times}
\usepackage{amssymb,amsmath}

\def\lsim{ \lower .75ex\hbox{$\sim$} \llap{\raise .27ex \hbox{$<$}} }
\def\gsim{ \lower .75ex \hbox{$\sim$} \llap{\raise .27ex \hbox{$>$}} }

\title[$\gamma$--ray  absorption and flat BLR in blazars] 
{``Flat" broad line region and $\gamma$--ray absorption in blazars}

\author[Tavecchio \& Ghisellini]
{F. Tavecchio$^1$\thanks{E--mail: fabrizio.tavecchio@brera.inaf.it} and
G. Ghisellini$^1$\\
$^1$INAF -- Osservatorio Astronomico di Brera, via E. Bianchi 46, I--23807
Merate, Italy\\
}

\begin{document}



\maketitle

\begin{abstract} 
We study the impact of the geometry of the broad line region (BLR) on 
the expected absorption, through the $\gamma\gamma\rightarrow e^{\pm}$ process,
of $\gamma$ rays produced in the relativistic jet of flat spectrum radio quasars (FSRQ). 
We consider ``flat" (or ``disky") BLR models, and use
BLR spectra calculated with the photoionization 
code {\tt CLOUDY}, already used to investigate the emission and the absorption 
of high--energy photons in FSRQ. 
We characterize the energy--dependent optical depth of the process, $\tau(E)$, 
for different accretion disk luminosities, aperture angles of the BLR 
($\alpha$, as measured from the equatorial plane), 
and initial injection eighths of the high-energy photons, $R_{\rm o}$.
We study in particular 
 how the change of these parameters influences the spectral break 
at GeV energies, predicted if the emission occurs within the BLR. 
We found a well defined relation between the break energy and  the post--break slope, 
both uniquely determined by $\alpha$. 
We finally find that even a rather disk--like BLR ($\alpha \sim 25^\circ$)
corresponds to important absorption ($\tau>1$) of photons above 
few tens of GeV produced within the BLR. 
We therefore conclude that the VHE emission detected from FSRQs occurs beyond the BLR.
%
\end{abstract}
 

\section{Introduction}

The presence of intense, broad ($v_{\rm FWHM}> 1000$ km s$^{-1}$) emission lines 
is one of the distinctive features of ``type--1" Active Galactic Nuclei 
(e.g. Osterbrock 1988, Netzer 2008). 
According to the current paradigm, lines originate in the so--called broad line region (BLR), 
where dense ($n=10^{10-11}$ cm$^{-3}$) clouds of gas, possibly part of a disk wind 
(e.g., Emmering et al. 1992, Elvis 2000, Trump et al. 2011), moving with keplerian 
velocity around the black hole (BH), are photo--ionized by the optical--UV emission 
of the central regions of the accretion disk. From the typical spectrum one can 
infer some of the basics properties of the clouds (temperature, density, composition; 
e.g., Kaspi \& Netzer 1999, Korista \& Goad 2000) and, using the so--called 
``reverberation mapping" technique (e.g., Peterson \& Wandel 2000), also their distance from the BH 
(e.g. Kaspi et al. 2000, 2007, Bentz et al. 2009) which can be used, in combination 
with the observed line width, to infer the mass of the BH (e.g. McLure and Dunlop 2001). 
Despite the decades of intense studies, some of the basic facts concerning the BLR are 
still under debate. 
In particular it is unclear whether the BLR is spherical or, instead, it is characterized 
by a ``flattened" geometry (e.g. Shields 1978), as expected in the case of a disk wind and 
as proposed for narrow--line Seyfert 1 galaxies, to reconcile the 
usually 
small inferred black hole masses with those of the broad line Sy1 (Decarli et al. 2008). 
Concerning the specific case of radio--loud AGN, several studies concluded that a flat 
geometry is favorite. 
For instance, early in the eighties Wills \& Browne (1986) found a correlation between 
the width of the $H\beta$ line and the core--to--extended radio flux at 5 GHz suggesting 
a BLR concentrated in the plane normal to the jet axis. 
Similar results have been more recently obtained by Jarvis \& McLure (2006). 
Decarli, Dotti \& Treves (2011) find the evidence that line width in blazars 
(in which, plausibly, the accretion disk is observed nearly on--axis) is smaller 
than in not aligned quasars, again suggesting that the BLR is flat, with an average 
thickness--radius ratio, $H/R\approx 0.05$.  

The bright $\gamma$--ray ($>100$ MeV) emission from blazars, in particular the 
powerful flat spectrum radio quasars (FSRQ), is widely interpreted as produced 
through to the inverse Compton scattering of soft photons by the relativistic 
electrons in the jet whose synchrotron continuum accounts for the low frequency 
(IR--optical) bump in the SED. 
The nature of the soft target photons is however still matter of debate: besides 
the synchrotron photons themselves (Maraschi et al. 1992) and the (``external") optical--UV photons from 
the broad line clouds (Sikora et al. 1994), possible other external sources include the UV--optical emission 
of the accretion disk (Dermer \& Schlickeiser 1993) and the IR thermal emission from the dusty torus (B{\l}a{\.z}ejowski et al. 2000). 
All these components are present at some level at all distances from the central black hole, 
but the relative importance changes with distance (e.g., Dermer et al. 2009, Sikora et al. 2009, 
Ghisellini \& Tavecchio 2009). 
In particular, photons from the BLR should be dominant at distances of $10^2$--$10^3$ Schwarzschild 
radii ($10^{16-17}$ cm for typical BH masses around $10^9$ M$_\odot$) commonly 
inferred from the variability timescale of these sources (e.g. Tavecchio et al. 2010, 
Foschini et al. 2010, Ackermann et al. 2010). 
Instead, IR photons or even internal synchrotron photons dominate if the emitting 
region is located far beyond the BLR, as argued by, e.g., Sikora et al. (2008), 
Marscher et al. (2008, 2010), Abdo et al. (2010). 
If the jet main emission region is located within the BLR, 
the emission line photons, besides acting as seeds for the IC process, are also
a source of {\it opacity} for the outgoing $\gamma$--ray emission at energies 
above 1 GeV through the reaction: $\gamma \gamma \rightarrow e^{\pm}$ 
(e.g. Donea \& Protheroe 2003, Liu \& Bai 2006, Tavecchio \& Mazin 2009, hereafter TM09, 
Poutanen \& Stern 2010, hereafter PS10). 
Assuming typical FSRQ parameters, the estimated optical depth $\tau$ 
above few tens of GeV would exceeds $\tau\sim 10$, making the detection 
of very high energy photons (VHE; $E>50 $ GeV) rather problematic in these sources.
The great majority of FSRQs, indeed, display rather steep GeV spectra, and opacity effects 
seem to explain quite well the ``universal" break observed at few GeV in the LAT spectra of 
FSRQs (PS10, Stern \& Poutanen 2011). 
However, the detection of a few FSRQs at VHE energies (Albert et al. 2008, Wagner et al. 2010, 
Aleksic et al. 2011a) implies a minimal degree of absorption, leading to propose that, 
at least during these events, the emission occurs {\it beyond} the BLR 
(Aleksic et al. 2011a,b, Tanaka et al. 2011, Tavecchio et al. 2011, Nalewajko et al. 2012, 
Dermer et al. 2012), although the observed fast variability requires a 
rather compact emission region. 

Previous works generally assumed a spherical geometry for the BLR. 
This paper aims at extending the study of the absorption of $\gamma$ rays in the BLR, already started in 
TM09, exploring in particular the consequences of a ``flat" geometry for the BLR. 
In this case one might expect a substantial reduction of the opacity to HE and VHE photons. 
This guess is based on the fact that the most effective collisions are those head--on, disfavored
in a flat geometry.  To this aim we calculate the expected BLR spectrum using the photoionizing code {\tt CLOUDY} 
(Ferland et al. 1998) with the assumption already used in Tavecchio \& Ghisellini (2008) 
and TM09, extending the calculation of $\tau$ to a flat geometry (\S2). 
Then we discuss the results, showing the optical depth for different combinations of the 
parameters specifying the system (\S3). 
We discuss the effect of such a geometry on the expected break at GeV energies discussed 
by PS10 and on the opacity at TeV energies. 
In \S4 we discuss the results.


\section{The model}

\begin{figure}
\hspace{-0.5cm}
\psfig{file=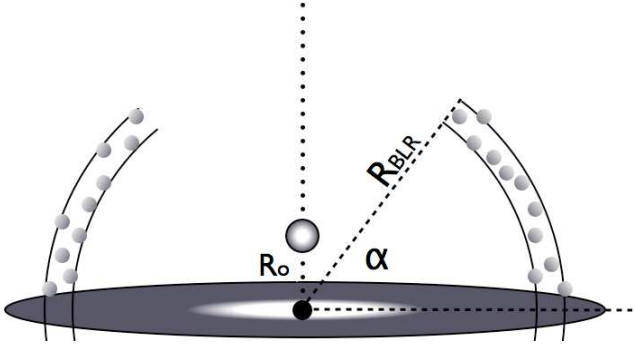,height=4.5cm}
\vskip 0.5 cm
\caption{
Sketch of the geometry for the calculation of the optical depth. 
$\gamma$--rays are produced in the jet at a height $R_o$ above the central black hole 
and travel into the radiation field of the BLR, characterized by aperture $\alpha$ and 
radius $R_{\rm \,\,  BLR}$. 
The accretion disc illuminates the clouds.}
\label{sketch}
\end{figure}

\subsection{BLR emission and geometry}

We refer to Tavecchio \& Ghisellini (2008) and TM09 for a deeper discussion of 
the model and of the assumptions used in the calculations. 
The assumed geometry is sketched in Fig. \ref{sketch}. 
Although there are indications pointing to a complex, stratified structure 
(e.g., Denney et al. 2009), for simplicity we model the BLR as a thin spherical shell 
with inner radius $R_{\rm BLR}$ filled with clouds characterized by density 
$n_{\rm gas}=10^{10}$ cm$^{-3}$ and column $N_{\rm H}=10^{23}$ cm$^{-2}$. 
To mimic ``flat" geometries we limit the BLR to an angle $\alpha$ (measured from the 
disc plane) and we assume that at larger angles there are no clouds (as in Fig. \ref{sketch}).  

\begin{figure}
\hspace{-0.3 truecm}
\psfig{file=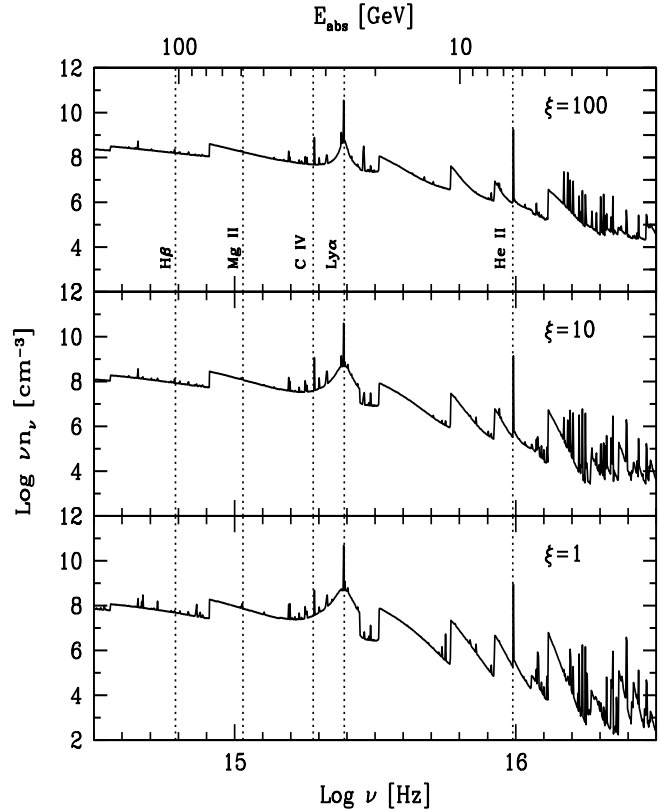,height=11.5cm,width=9.5cm}
\vskip -0.3 cm
\caption{
Zoom on the optical--UV band of BLR spectrum for different values of the ionization 
parameter, $\xi=10$, $10^2$ and $10^3$. 
Dotted vertical lines indicate the position of the most important emission lines
(as labeled).
The two most prominent lines are the Ly$\alpha$ lines of H and He II. 
These are also the two lines providing most of the opacity at GeV energies. 
The top $x$--axis reports the energy of the target $\gamma$--ray photons, calculated 
for the maximum of the cross section and head--on collisions (see text).}
\label{lines}
\end{figure}

\begin{figure*}
\psfig{file=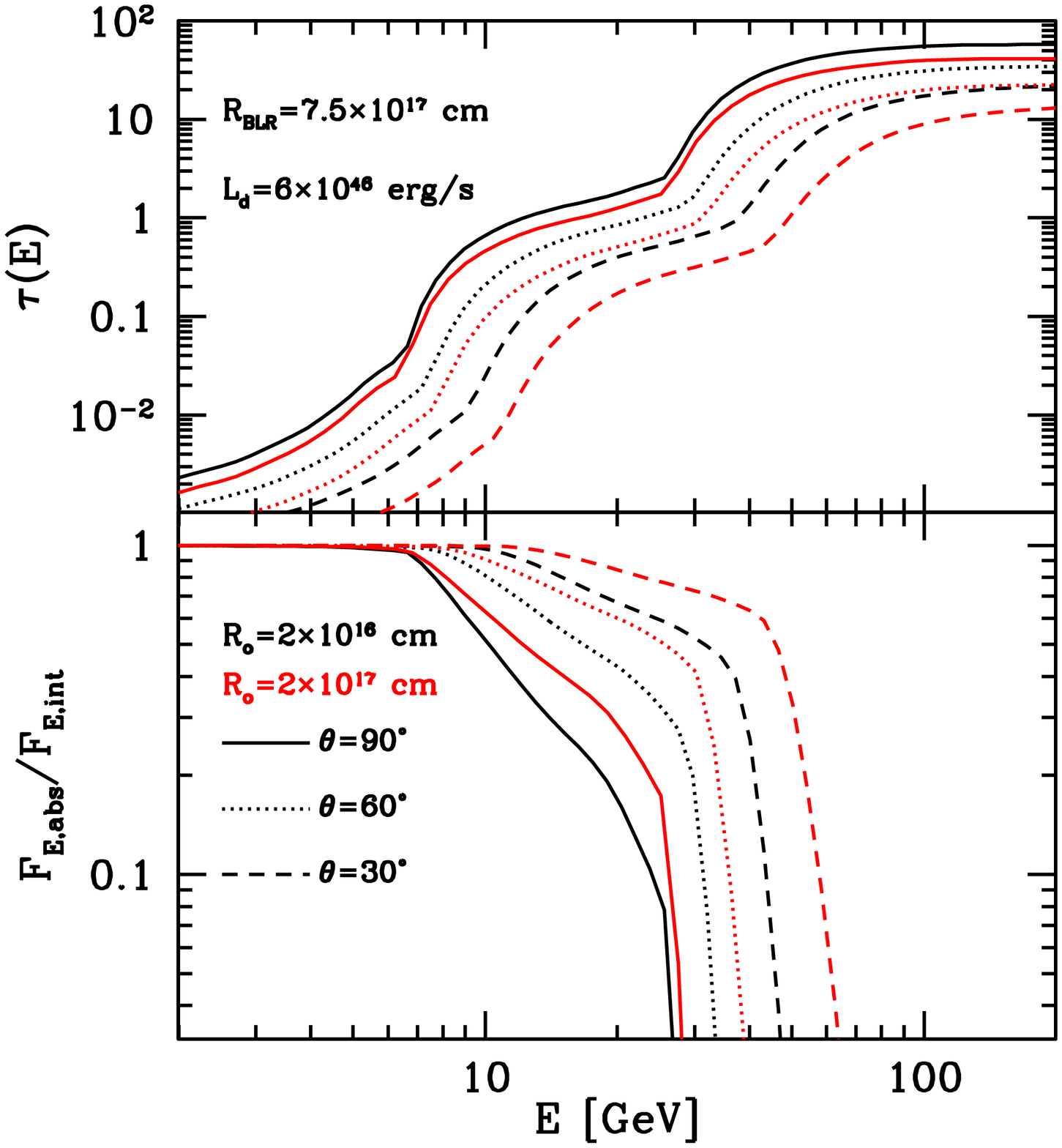,height=8.8cm,width=8.8cm}
\vspace{-8.8 cm}
\hspace{8 cm}
\psfig{file=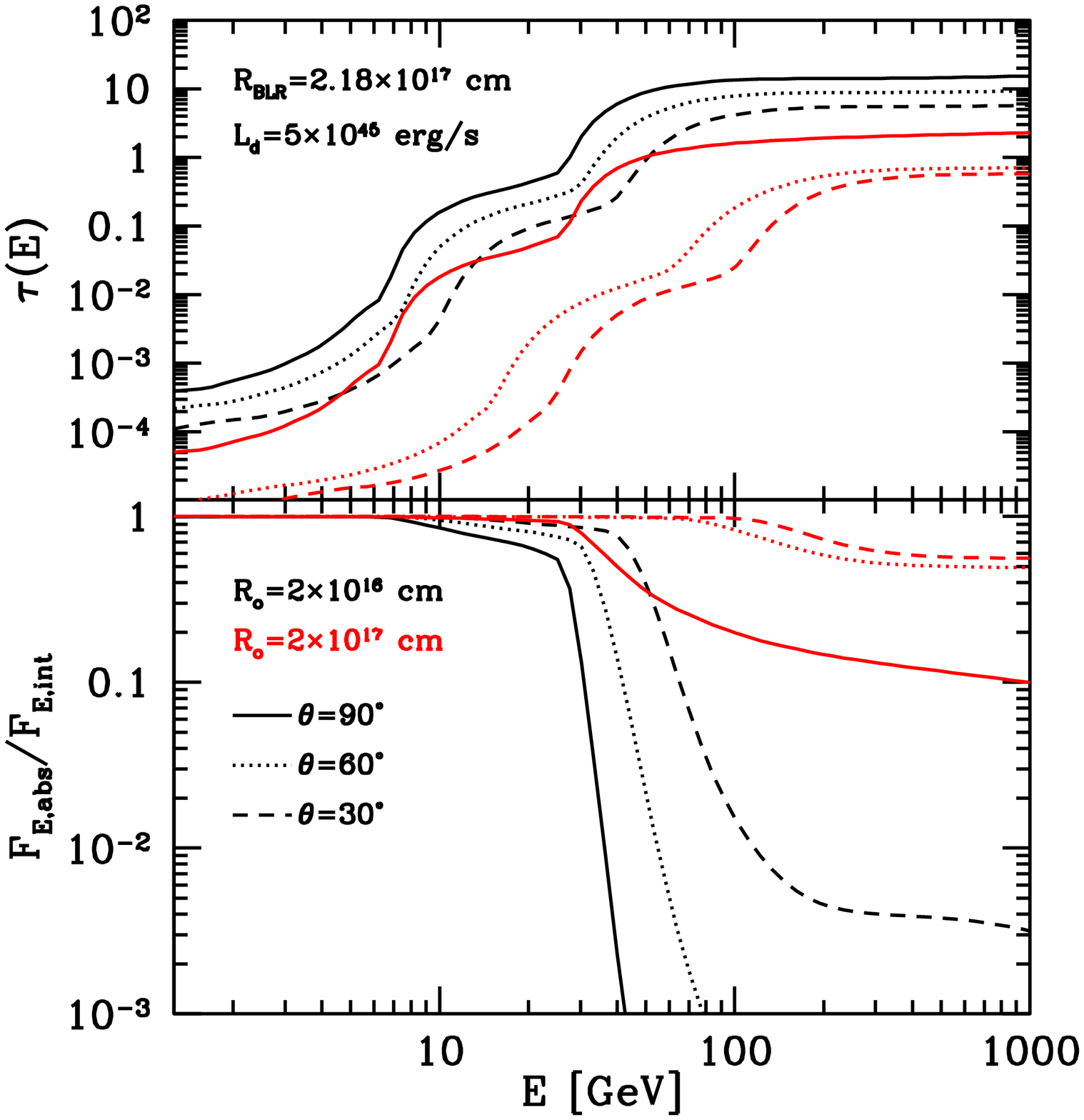,height=8.8cm,width=8.8cm}
\caption{
{\it Upper panel}: optical depth of $\gamma$ rays as a function of the energy 
for different parameters. 
Solid, dotted and dashed lines report the results for BLR angles 
$\alpha=90^\circ$, 60$^\circ$ and 30$^\circ$. 
Black and red lines are for initial distances $R_{\rm \, o}=2\times 10^{16}$ cm and 
$R_{\rm \, o}=2\times 10^{17}$ cm, respectively. 
{\it Lower panel}: Absorbed over intrinsic flux ratio. 
Parameters as above. Note the spectral breaks in correspondence to ``jumps" in the optical depth.
The left plot has been calculated for a disk luminosity $L_{\rm d}=6\times 10^{46}$ erg s$^{-1}$. 
The right panel is for $L_{\rm d}=5\times 10^{45}$ erg s$^{-1}$. 
In both cases $R_{\rm \,BLR}$ is calculated following the scaling 
adopted by Ghisellini \& Tavecchio (2009).}
\label{tau}
\end{figure*}

Based on the typical line/continuum luminosity ratio we always assume that BLR clouds 
intercept a fraction $C=\Omega _{BLR}/2\pi = 0.1$ of the illuminating continuum 
(we consider only the emisphere toward the observer, since the other one is occulted by the disk). 
In the case of a ``flat geometry" this requirement implies a {\it lower} limit to 
the BLR angle, $\alpha_{\rm min}$. The existence of this limit  can be simply understood considering that, in order to 
keep $C$ constant for decreasing $\alpha$, the clouds are forced to occupy a decreasing 
volume, until they will completely fill all the available space. 
In that case the solid angle of the geometrical structure containing the clouds 
satisfies $\Omega_{\rm BLR}=2 \pi C$, that is:
\begin{equation}
\int_{0}^{\alpha _{\rm min}} 2\pi \sin\alpha \, d\alpha = 2\pi C  
\,\, \rightarrow \,\, \alpha_{\rm min}=\arccos (1-C)
\label{alphamin}
\end{equation}
which, for $C=0.1$, gives $\alpha_{\rm min}\simeq 25^\circ$. For simplicity, 
in this calculation (and in the entire paper) an {\it isotropic} 
photo--ionizing radiation field is assumed. 
For a geometrically thin accretion disk, for which the flux depends on the 
cosine of the angle from the jet axis (e.g., Frank, King \& Raine 2002), one would obtain a larger value 
of $\alpha_{\rm min}$, due to the depression of the disk flux at large angles. Given this limit, in the following example we assume $\alpha>25^\circ$.

The central illuminating continuum is modelled as a combination of an UV bump 
and a flat X--ray power law  (see Korista \& Goad 2001 and Tavecchio \& Ghisellini 2008) 
and total luminosity $L_{\rm d}$. We assume a disk temperature $T_{\rm d}=1.5\times 10^5$ K. 


The main features of the spectrum produced by BLR clouds depend on the 
value of the ionization parameter, $\xi$. 
We define $\xi \equiv L_{\rm d}/n_{\rm gas}R^2_{\rm BLR}$, using the bolometric 
luminosity of the ionizing continuum. 
Fig. \ref{lines} shows three examples of spectra in the optical--UV 
region [represented as $\nu n(\nu)$, the relevant quantity for absorption, where $n(\nu)$ 
is the specific photon energy density] for three different values of $\xi=10, 10^2$ and $10^3$. 
While some of the features change with $\xi$, the most prominent lines, 
such as the hydrogen ($\nu =2.5\times 10^{15}$ Hz) and the helium II 
($\nu =10^{16}$ Hz) Ly$\alpha$ lines, are rather stable (see also PS10). 

To reduce the number of free parameters, for a given disk luminosity 
$L_{\rm d}$ we have fixed the BLR radius according to the relation 
\begin{equation}
R_{\rm BLR}=10^{17} \left( \frac{L_{\rm d}}{10^{45} {\rm erg \, s}^{-1}} \right)^{0.5} \, {\rm cm} 
\label{rblr}
\end{equation}
(e.g., Ghisellini \& Tavecchio 2009) which is a good approximation of 
the results of the reverberation mapping technique. 
Therefore, since $n\propto L_{\rm d}/R^2_{\rm BLR}$, 
$\tau\propto n R_{\rm BLR} \propto L_{\rm d}^{0.5}$ increases with the 
disk luminosity while the ionization parameter $\xi=10$ is constant 
(corresponding to the spectrum shown in the middle panel of Fig. \ref{lines}). 
As a consequence, the ``shape" of the optical depth as a function of the photon energy 
is fixed, and only its normalization changes with the disk luminosity.

\subsection{Absorption of $\gamma$ rays}

Gamma--ray photons are produced in a region (assumed stationary) located at 
an eighth $R_{\rm o}$ over the central BH and propagate within the radiation field of the BLR. 
The optical depth for the $\gamma \gamma \rightarrow e^{\pm}$ process, 
$\tau (E)$, is calculated as (e.g., Liu \& Bai 2006):
\begin{equation}
\tau(E)=\int_{R_o}^{\infty} dx\int d\Omega \int d\nu \, 
n(\nu,\Omega ,x) \sigma _{\gamma \gamma}(E,\nu,\Omega) (1-\mu )
\label{eqtau}
\end{equation}
where $E$ is the energy of the $\gamma$--rays, $x$ is the distance of the photons 
from the BH, $\mu=\cos \theta$, where $\theta $ is the collision angle, 
$n(\nu,\Omega,x)$ is the angle--dependent number density of the BLR photons at 
each location of the photon path, $l$, $\sigma _{\gamma\gamma}(E,\nu,\Omega)$ 
is the pair production cross section (e.g., Heitler 1960) and $d\Omega =-2\pi d\mu$ 
is the solid angle covered by each portion of the BLR at a given position $x$.  For 
simplicity of calculation, the upper limit of the integral over $x$ has been set 
to $5 R_{\rm BLR}$.

\begin{figure*}
\vskip -0.8 cm
\psfig{file=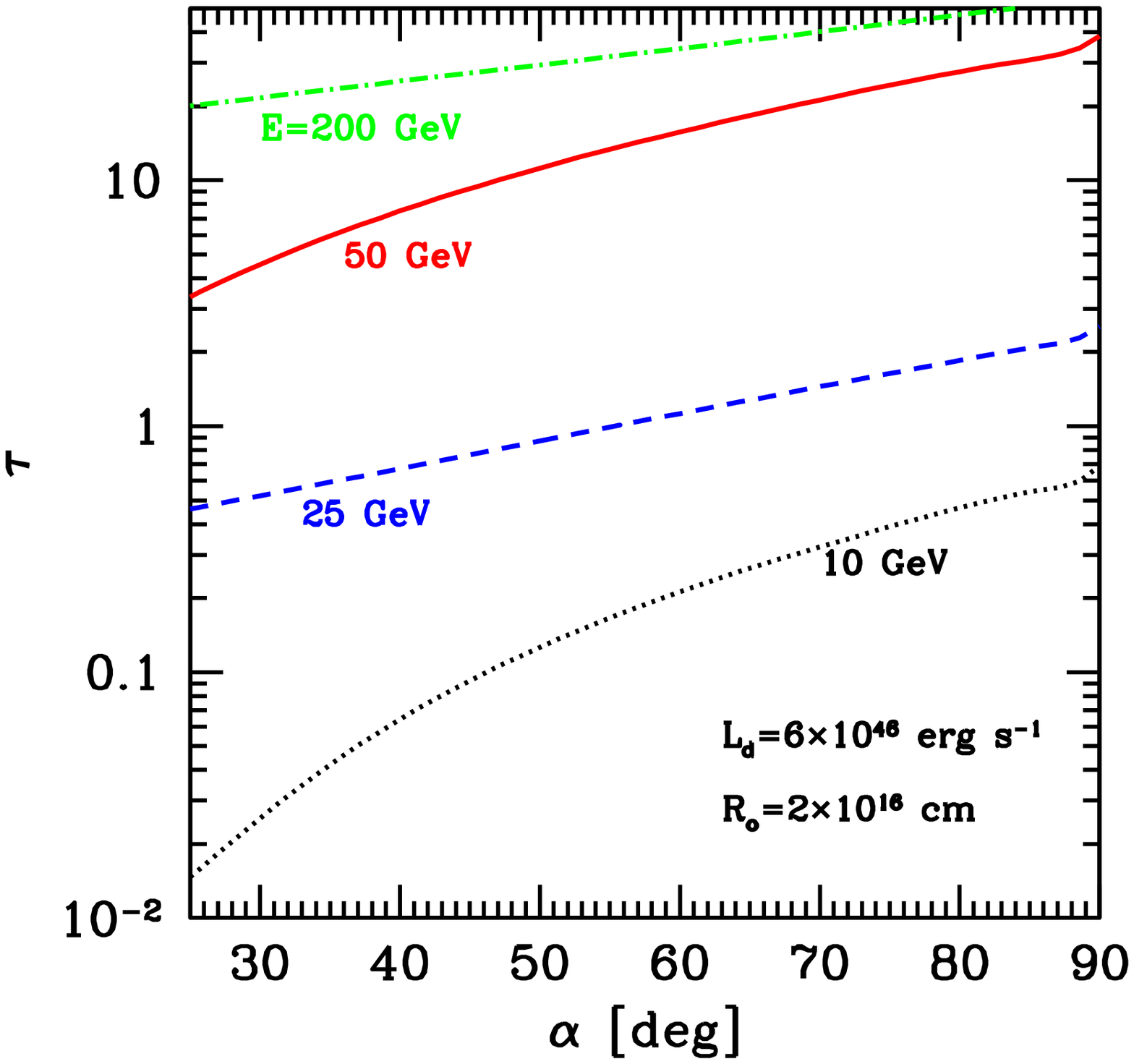,height=9.5cm,width=9.5cm}
\vspace{-9.5 cm}
\hspace{7.8 cm}
\psfig{file=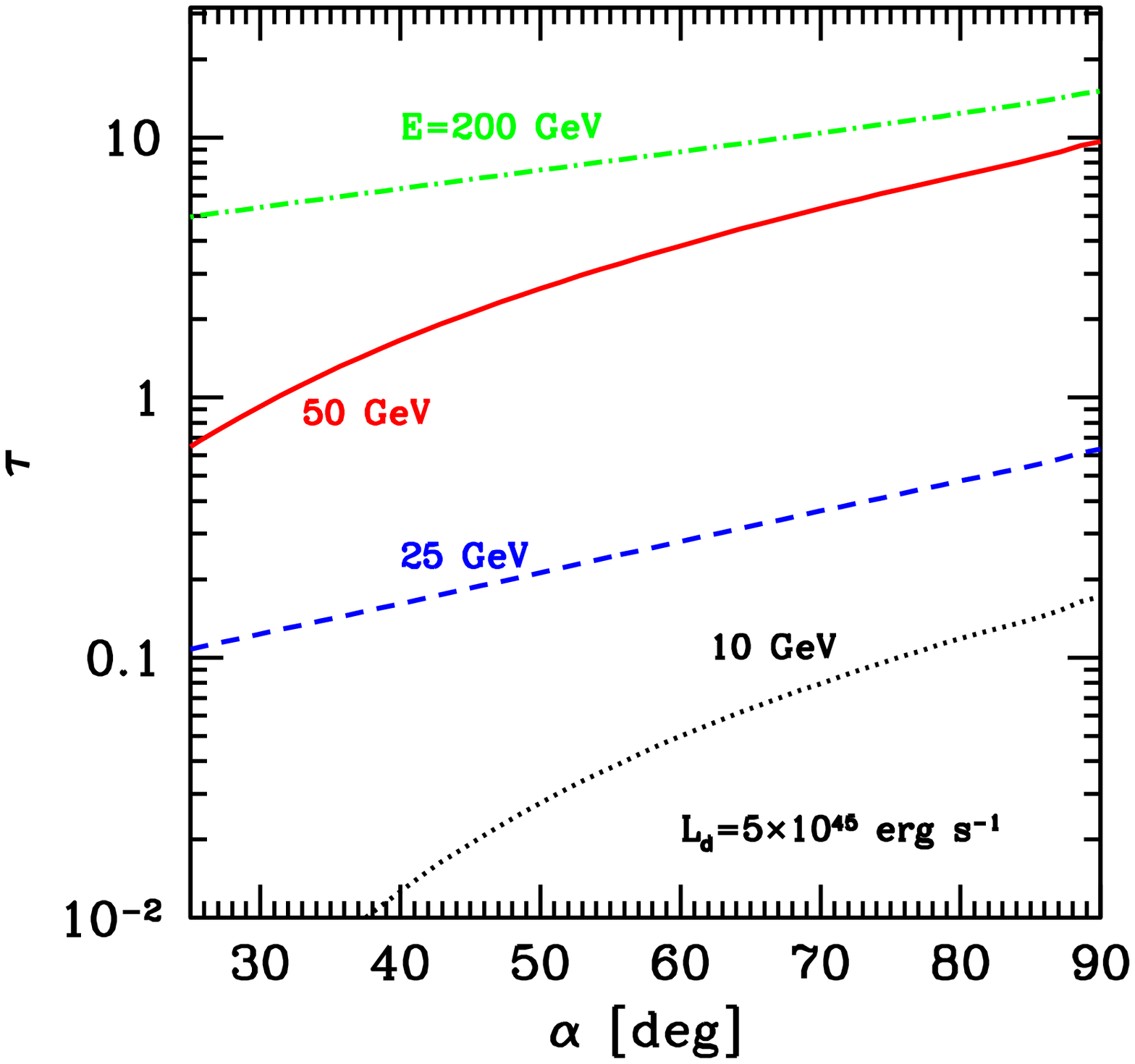,height=9.5cm,width=9.5cm}
\vskip -0.5 cm
\caption{
Dependence of $\tau$ on the aperture angle $\alpha$ of the BLR for different values 
of the photon energies: $E=10$ GeV (dotted, black), 25 GeV (dashed, blue), 50 GeV (solid, red) 
and 200 GeV (dot--dashed, green). 
Left panel: $L_{\rm d}=6\times 10^{46}$ erg  s$^{-1}$. 
Right panel: $L_{\rm d}=5\times 10^{45}$ erg s$^{-1}$. 
In both cases the distance of the emitting region from the central black hole 
has been fixed to $R_{\rm \, o}=2\times 10^{16}$ cm.
}
\label{taualpha}
\end{figure*}

Thanks to the existence of a sharp kinematic threshold for the process and of a 
well defined maximum of the cross section, for a given frequency of the 
target photons it is possible to associate the energy of the $\gamma$ rays 
most effectively interacting and absorbed (and {\it viceversa}). From the threshold definition we have:
\begin{equation}
h\nu = \frac{2 m_e^2 c^4}{(1-\mu) E}
\label{thre}
\end{equation}
Considering head--on ($\mu=-1$) collisions one obtains the threshold 
$\nu \approx 6\times 10^{14} (E/100 \, {\rm GeV})^{-1}$ Hz. 
The top $x$--axis of Fig. \ref{lines} reports the energy $E$ calculated with this relation. 
Given this strict link between $\nu$ and $E$, the optical depth $\tau(E)$ 
reflects the spectrum of the soft photon field $n(\nu)$ (smeared by the cross section). 
In particular (as pointed out by PS10), the H and HeII Ly$\alpha$ line will 
imprint particular features in the spectrum at typical  rest frame energies of $\approx 25$ GeV and 6 GeV.

Due to the kinematics of the process, the most effective collisions are 
those head--on. Therefore a ``closed" (i.e., $\alpha =\pi/2$) BLR will be more opaque 
than an ``open" one for which head--on collisions are suppressed. 
Moreover, since the threshold for absorption is angle dependent, also the 
characteristic relation between the $\gamma$--ray energy and the corresponding
frequency of the target will change (note the angular term in Eq. \ref{thre}). 
This effect implies a change of the threshold energy when changing $\alpha$ and $R_{\rm o}$.

\section{Results}

In our scheme, the optical depth $\tau(E)$ depends on the disk luminosity 
$L_{\rm d}$ and on two geometrical parameters: the BLR angle $\alpha$ and 
the eighth of the injection region, $R_{\rm o}$.

Concerning the luminosity, in the following examples we focus on two values, 
$L_{\rm d}=6\times10^{46}$ erg s$^{-1}$  and $5\times10^{45}$ erg s$^{-1}$. 
These particular values bracket the distribution of disk luminosities of $\gamma$--ray 
emitting FSRQ (e.g., Ghisellini et al. 2010). 
Moreover, they characterize some  ``prototypical" FSRQs: 
the higher luminosity is appropriate for the accretion disk in 3C454.3 
(e.g., Bonnoli et al. 2011), 
while the lower value is close to the luminosity of the disk in 3C279 (Pian et al. 1999).  
The corresponding radii of the BLR, estimated with the scaling provided 
by Eq. \ref{rblr}, are $R_{\rm BLR}=7.5\times 10^{17}$ cm and $2.18\times 10^{17}$ cm.

\begin{figure*}
\vskip -0.7 cm
\psfig{file=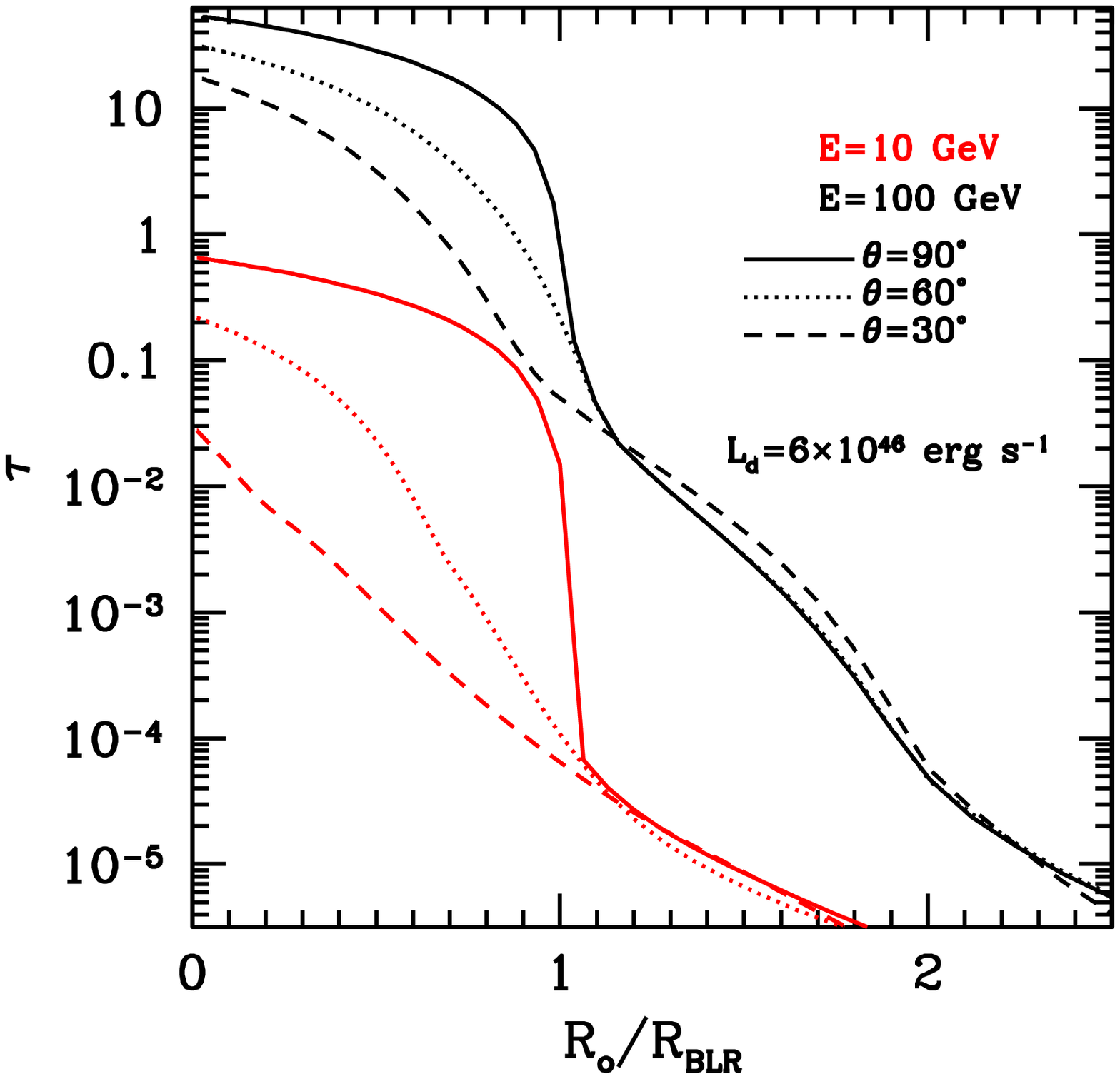,height=9.5cm,width=9.5cm}
\vspace{-9.55 cm}
\hspace{7.5 cm}
\psfig{file=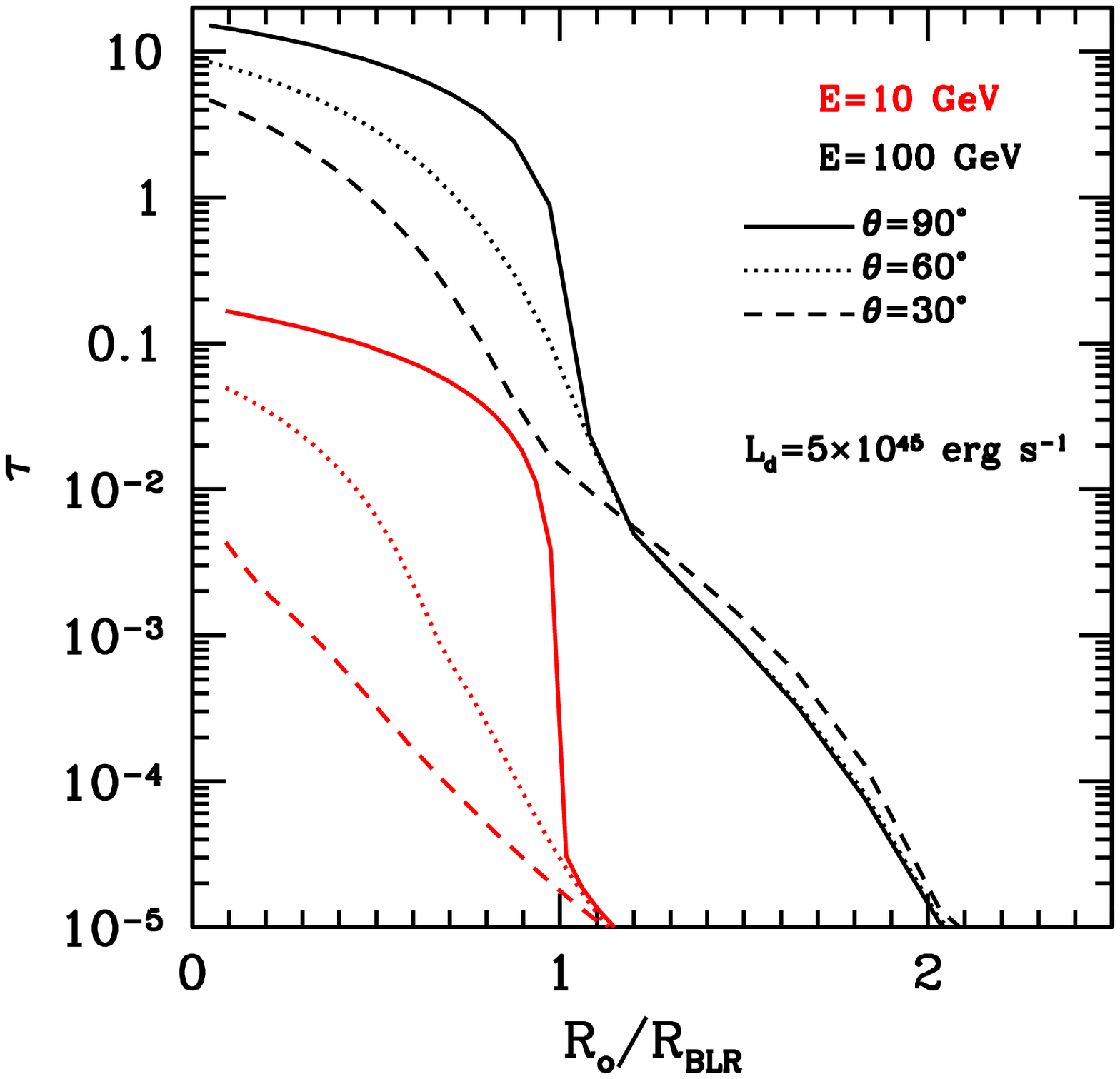,height=9.5cm,width=9.5cm}
\vskip -0.5 cm
\caption{
Dependence of $\tau$ on the position of the injection of $\gamma$ rays for 
different values of the photon energy (10 GeV, red and 100 GeV black) and BLR angle (solid line: $\alpha=90^\circ$; 
dotted: 60$^\circ$; dahsed: 30$^\circ$). 
The right and left panels show the case with $L_{\rm d}=6\times 10^{46}$ erg s$^{-1}$ 
and $L_{\rm d}=5\times 10^{45}$ erg s$^{-1}$, respectively.}
\label{taudist}
\end{figure*}

Having fixed the luminosities, the model is now only specified by the 
geometrical parameters $\alpha$ and $R_{\rm o}$.
Fig. \ref{tau} (upper panels) shows the optical depth as a function 
of energy, $\tau(E)$, calculated with our model for different values of the 
BLR angle $\alpha$ and initial distance $R_o$ and for the two different 
values of $L_{\rm d}$.
Fig. \ref{taualpha} and Fig. \ref{taudist} show $\tau$ 
as a function of $R_o$ and $\alpha$ for different values of the photon energy.

All the curves in Fig. \ref{tau} have the same structure, strictly linked 
to the features of BLR spectrum. 
In all cases the optical depth monotonically increases with energy and two well 
defined ``jumps" are visible, where the optical depth suddenly increases. 
These two steps mark the threshold energy for collisions with target
photons belonging to the HeII and H Ly$\alpha$ lines (see the discussion in 
PS10). For a closed geometry (solid lines) the critical energies 
reflect the threshold energies for the most effective head--on ($\theta=\pi$) 
collisions, $E_{\rm He II}= 6$ GeV and $E_{\rm H}= 25$ GeV using 
Eq. \ref{thre} (more discussion in \S 3.1).

As already noted, the value of the energy threshold is angle dependent. 
Therefore the position of the jumps in $\tau(E)$ depend on the BLR geometry 
and position of the emission region. 
In particular, for opened geometries, the corresponding curves preserve their shape 
but they are approximately shifted toward higher energies and the overall level decreases. 
Both effects are easy to understand: the general decrease of $\tau$ is in fact 
expected because of the reduced efficiency (the typical interaction angle 
progressively decreases), while the shift in energy is due to the $(1-\mu)$ 
dependence in the threshold (Eq. \ref{thre}): target photons of a given energy 
interact with $\gamma$ rays with increased energy. 
This effects are clearly visible in Fig. \ref{tau} where, together with the case 
of the ``closed" geometry, we also report the results for $\alpha=60^\circ$ and 30$^\circ$. 

The same kinematic and angular effects also regulate the behavior of the 
curves in Fig. \ref{taualpha} and \ref{taudist}.
In particular, the behavior of the curves in Fig. \ref{taudist}, showing 
the dependence of the opacity on the injection heigth, deserve some comments. 
In the case of the ``closed" geometry ($\alpha=\pi/2$, solid curves), the optical 
depth gradually decreases for increasing values of $R_{\rm o}$ as long as 
$R_{\rm o}<R_{\rm BLR}$ since $\gamma$--ray photons travel for progressively 
smaller distances within the BLR. 
When the injection heigth coincides with the BLR radius, the optical depth 
shows a rapid decrease, caused by the abrupt transition from a situation 
in which $\tau$ is dominated by head--on ($\theta = \pi$) collision to 
that in which only unfavorable collision angles ($\theta < \pi/2$) are allowed. 
The same behaviour is followed by the curves corresponding to different 
$\alpha$ (dotted lines: 60$^\circ$; dashed: 30$^\circ$) although the 
decrease of $\tau$ at $R_{\rm o}=R_{\rm BLR}$ is smoother  due to the 
intrinsic absence of the critical head--on collisions.


\subsection{Effect on the spectrum: the ``GeV break"}

All the features discussed above for the optical depth $\tau(E)$ are reflected by the 
out--going {\it absorbed} $\gamma$--ray spectrum of FSRQ. 
In particular, as pointed out in PS10, for power law 
intrinsic spectra $F_{\rm int}(E)\propto E^{-\Gamma}$, the ``jumps" in the 
optical depth imprint in the absorbed spectrum (lower panels in Fig. \ref{tau}), 
spectral {\it breaks} where the observed spectrum displays a sudden change 
of slope, i.e. $F_{\rm obs}(E)=F_{\rm int}(E) \, e^{-\tau(E)}
\propto E^{-(\Gamma + \Delta \Gamma)}$ above the break energy 
$E>E_{\rm br}$ corresponding to the energy threshold for a given emission line. 
The breaks are clearly visible in the cases in which the optical depth approaches unity, $\tau\gtrsim 0.1$.
For a closed geometry, these breaks are located always at the same energy and 
PS10 argued that this effect associated to the HeII line 
could account for the ``universal" energy break observed at few GeV in the LAT spectra of FSRQs.  

\begin{figure}
\vskip -0.8 cm
\psfig{file=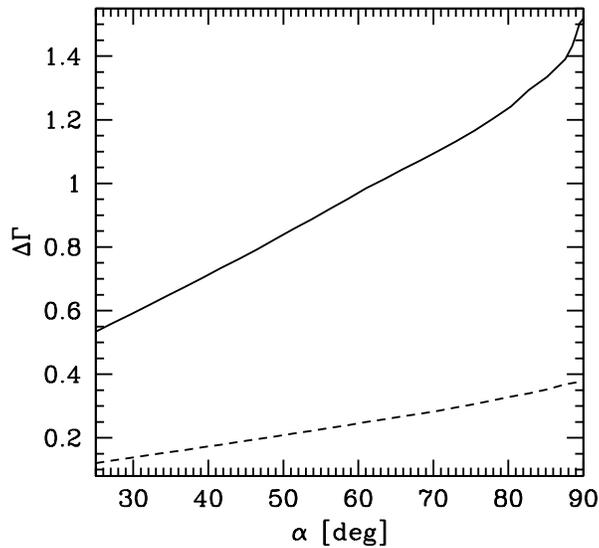,height=8.5cm,width=8.5cm}
\caption{
Change of the spectral slope $\Delta \Gamma$ at the HeII threshold in the 
absorbed $\gamma$--ray spectrum as a function of the BLR angle $\alpha$ for 
$L_{\rm d}=6\times 10^{46}$ erg s$^{-1}$ (solid) and $L_{\rm d}=5\times 10^{45}$ 
erg s$^{-1}$ (dashed).}
\label{deltagamma}
\end{figure}

\begin{figure}
\vskip -0.6 cm
\psfig{file=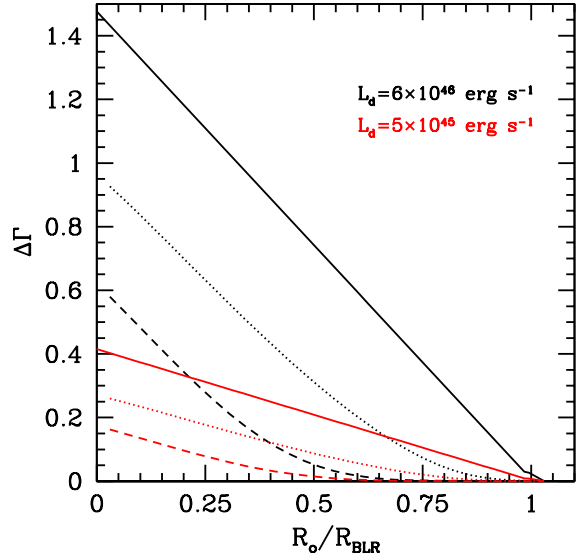,height=8.5cm,width=8.5cm}
\caption{
Change of the spectral slope $\Delta \Gamma$ at the HeII threshold 
in the absorbed gamma-ray spectrum as a function of the emission eighth $R_{\rm o}$ 
for $\alpha=90^\circ$ (solid), 60$^\circ$ (dotted) and 30$^\circ$ (dashed). 
Black curves: $L_{\rm d}=6\times 10^{46}$ erg s$^{-1}$. 
Red curves: $L_{\rm d}=5\times 10^{45}$ erg s$^{-1}$.}
\label{deltagammaro}
\end{figure}

\begin{figure}
\vskip -0.4 cm
\psfig{file=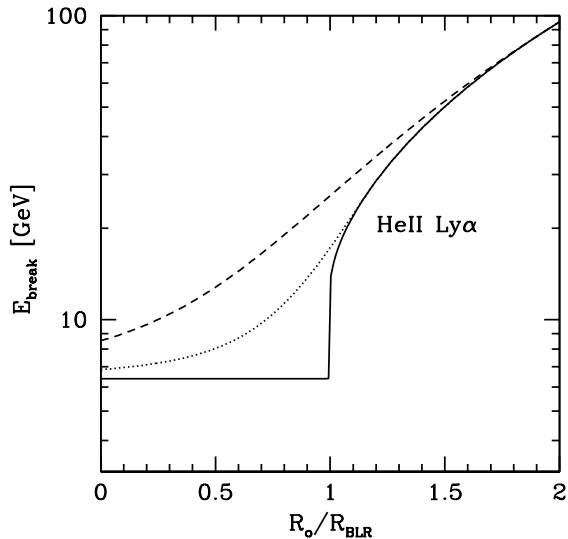,height=7.8cm,width=8cm}
\caption{
Break energy at the HeII threshold in the absorbed $\gamma$--ray spectrum 
as a function of the emission eighth $R_{\rm o}$ for $\alpha=90^\circ$  (solid), 
60$^\circ$ (dotted) and 30$^\circ$ (dashed).}
\label{break}
\end{figure}

The variations of the energy of the jumps in $\tau$ caused by changes in 
$\alpha$ and $R_{\rm o}$ translate in the change of the break energies in the observed spectrum. 
Furthermore, since the change of the slope at the break is proportional 
to the optical depth around $E_{\rm br}$, $\Delta \Gamma \propto \tau(E_{\rm br})$ (PS10), 
not only the energy of the break, but also $\Delta \Gamma$ will depend on $\alpha $ and $R_{\rm o}$. 
Both effects are clearly visible in the lower panel of Fig. \ref{tau}. 
Figs. \ref{deltagamma}--\ref{break} separately show the dependence of $\Delta \Gamma$ 
and  $E_{\rm br}$  on the geometrical parameters. 
We limit the results to the case of the HeII break whose energy is the most 
accessible with {\it Fermi}/LAT data. 
As expected, $\Delta \Gamma$, directly proportional to the optical depth, decreases 
with decreasing $\alpha$ and increasing $R_{\rm o}$. 
On the other hand, the dependence of $E_{\rm br}$ on $R_{\rm o}$, due to the angular 
dependence of the threshold energy in Eq. \ref{thre}, is more complex. For $\alpha=\pi/2$ 
the opacity is mostly determined by head--on collision as long as $R_{\rm o}< R_{\rm BLR}$ 
and thus HeII line photons will produce the break at $E_{\rm HeII}$. For emission sites 
beyond the BLR, instead, soft photons are coming at angles $\theta <\pi/2$ and thus, 
according to the angular dependence in Eq. \ref{thre}, the break will move to higher energies. 

For opened geometries, instead (with no head--on collisions), the increase of $R_{\rm o}$ determines
a decrease of the typical collision angle between $\gamma$ rays and BLR photons which, due to the angular dependence of the threshold energy, translates in the  monotonic increase of $E_{\rm br}$ with $R_{\rm o}$. 
Both dependences reflect that of $\tau$ in Figs. \ref{taualpha} and \ref{taudist}. 
Evidence for variations of the optical depth $\tau(E_{\rm br})$ (and correspondingly of 
$\Delta \Gamma$) has been found by Stern \& Poutanen (2011) analyzing the 
{\it Fermi}/LAT spectra of the bright FSRQ 3C454.3 at different epochs. 
They interpret these variations as due to different location of the emitting region in a {\it stratified} BLR, in which (high--ionization) HeII 
lines are produced at radii smaller than low--ionization lines. 
Emission episodes occurring well within the BLR, close to the ``HeII BLR" would result in 
strong absorption and large $\Delta \Gamma$. 
If $\gamma$--rays are instead produced at larger radii, the absorption through HeII photons is 
suppressed and, correspondingly, the break around the HeII threshold energy is less pronounced. 

\begin{figure}
\vskip -0.4 cm
\psfig{file=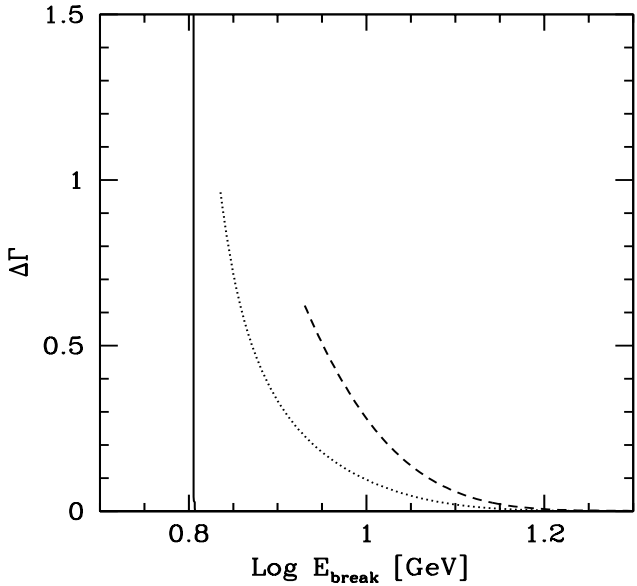,height=8.5cm,width=9.3cm}
\caption{
Change of the spectral slope $\Delta \Gamma$ at the HeII threshold in the absorbed 
$\gamma$--ray spectrum as a function of the break energy in the spectrum for the 
case of the emission inside the BLR (i.e. $R_{\rm o}<R_{\rm BLR}$) for 
$\alpha=90^\circ$  (solid), 60$^\circ$ (dotted) and 30$^\circ$ (dashed). For $\alpha=90^\circ$ 
the break energy is fixed since absorption is dominated by head--on collisions.
}
\label{dgvseb}
\end{figure}

The flat BLR scenario allows a different, although related, interpretation. 
Also in this case, in fact, the opacity depends on the distance of the emission region 
from the central black hole. 
This leads to envisage scenarios in which, analogously to the model of Stern \& Poutanen (2011), 
the location of the emission region in the jet changes with time, causing the opacity to 
$\gamma$--rays to change. 
In this case, however, the variations are fundamentally linked to the angular dependence 
of kinematics and energy threshold of the absorption process.  
If the BLR is ``closed" the break energy do not change as long as the emission occurs 
within the BLR but $\Delta \Gamma$ decreases for increasing $R_{\rm o}$ (Fig. \ref{deltagammaro}). 
Instead, for ``opened" geometries, both $\Delta \Gamma$ and $E_{\rm br}$ depend on $R_{\rm o}$ 
(Fig. \ref{deltagammaro} and \ref{break}). 
In this case, see Fig. \ref{dgvseb}, a well defined connection between 
$\Delta \Gamma$ and $E_{\rm b}$ is expected.  

The dependence of $\Delta \Gamma$ and $E_{\rm br}$ with geometry, summarized in 
Figs. \ref{deltagammaro}--\ref{break}, allows us to envisage several interesting cases. For 
instance, a break at $E_{\rm br}>E_{\rm HeII}$ surely points to an open geometry. 
If, in addition, $E_{\rm br}$ and $\Delta \Gamma$ do not change from state to state, we 
can conclude that the emitting source is approximately located at a fixed distance from the black 
hole (as expected in the case of a standing shock). 
Alternatively, a correlated variation of $E_{\rm br}$ and $\Delta \Gamma$ would allows us, 
in principle, to determine $\alpha$.  
These examples show that the flattened BLR scenario is clearly testable and predictive, 
allowing, for instance, to determine $\alpha$ if correlated variations of $\Delta \Gamma$ 
and $E_{\rm br}$ are observed.

\subsection{Opacity at VHE}

As discussed in the Introduction, one issue that motivated our study was to explore the possibility to avoid important 
absorption at high energy ($E>50$ GeV). 
Indeed, as already extensively discussed in literature 
(e.g. Liu \& Bai 2006, Sitarek \& Bednarek 2008, TM09, PS10), 
under standard assumptions the BLR is expected to be strongly opaque above few tens of GeV. 
This fact forces to conclude that the VHE emission detected from FSRQs is produced outside 
the BLR (e.g., Tanaka et al. 2011, Aleksic et al. 2011a,b, Tavecchio et al. 2011, 
Dermer et al. 2012), even if the short variability timescale points to rather compact 
regions (for a non--standard view considering photon--axion conversions, see Tavecchio et al. 2012).

Fig. \ref{tau} shows that for $\alpha=\pi/2$  (``closed" BLR) the optical depth exceeds 
10 above 30 GeV for both low and high luminosity if the emission region is 
located well inside the BLR. For high 
luminosities the same is true even considering the case of open BLR with $\alpha$ as small as 30$^\circ$ 
(left panel). 
At low luminosities, because of the $\tau\propto L_{\rm d}^{1/2}$ dependence, 
the optical depth is generally smaller than in the high luminosity case but it is 
still much larger than 1 for a standard BLR and a jet emission region located inside the BLR. 
Only for flat geometries and large distances $R_{\rm o} (\gtrsim R_{\rm BLR}$), $\tau$ decreases 
below a few. 
This can be better appreciated Fig. \ref{taualpha} and Fig. \ref{taudist}, reporting 
the dependence of $\tau$ with $\alpha$ and $R_{\rm o}$. 
Clearly, for energies above 100 GeV the absorption is important for almost all values of $\alpha$ 
if $R_{\rm o}< R_{\rm BLR}$. 
Again, $\tau<1$ only if $R_{\rm o} \gtrsim R_{\rm BLR}$.

Concluding,  we showed that for all the possible geometries and luminosities it is very hard to avoid very large
absorption even in the case of flattened geometries.  
Thus FSRQ are expected to be strongly opaque at high energy photons, $E>50$ GeV if the emission 
occurs well within the BLR.  Therefore, as in Tavecchio et al. (2011), for the case of the 
VHE flare of PKS 1222+200 (and also for 3C279, Aleksic et al. 2011b), one is forced 
to assume that the emitting region is located at large distances ($R_{\rm o}>R_{\rm BLR}$) even 
in the case of extreme flat geometries.

\section{Conclusions}

We have studied the absorption of $\gamma$ rays through the interaction with the 
UV--optical photons emitted by the BLR in FSRQ assuming a ``flattened" geometry. 
We showed that the geometry has an important impact in the resulting opacity to 
$\gamma$ rays, that can be traced back to the angular dependence of the kinematics 
and the pair production energy threshold. 
In particular we have discussed the implication of the BLR geometry for the ``GeV break" 
imprinted in the $\gamma$--ray spectra at energies corresponding to the most prominent emission lines. 
While in the case of a completely spherical BLR the energy of these spectral features 
is fixed, corresponding to the threshold energy associated to lines photons interacting 
head--on with $\gamma$ rays, an opened geometry implies a dependence of $E_{\rm br}$ 
on the location of the jet emission site. 
Together with the parallel dependence of $\Delta \Gamma$, the change of spectral 
index at the break, this dependence could in principle be used to reconstruct  
the BLR geometry and the location of the emission region.

Of course several aspects of our treatment are highly idealized. 
The possible stratification of the BLR, with different emission lines produced at 
different distances from the photo--ionizing source has been neglected (e.g., Krolik 1999, Stern \& Poutanen 2011). 
We also do not consider the possibility that electrons associated to a tenuous warm gas filling the inner AGN regions 
can partly scatter and isotropize the BLR and accretion disc photons, making the angular dependence at 
the base of the effects less sharp. Finally, we do not include in the calculation the IR radiation field emitted by the 
hot ($T\approx 500-1000$ K) dusty torus, which is expected to be an important source of opacity above few hundreds of GeV (e.g., Donea \& Protheroe 2003).

Finally, we mention that another issue worth to be explored is the impact of the BLR 
geometry on the high--energy emission produced through the inverse Compton scattering of the 
BLR photons, envisaged in the external Compton models (e.g., Tavecchio \& Ghisellini 2008). 
We can expect that an opened geometry could lead to changes in, e.g., typical frequencies 
and luminosities of this component, thought to dominate the $\gamma$--ray emission of FSRQs. These changes, in turn, would translate in different inferred parameters for the emissione zone.

Besides these issues, that can be explored in a future work, the clear--cut results that we derived make the scheme attractively predictive and testable, and could provide an effective probe of the $\gamma$--ray emission sites of FSRQ.

\section*{Acknowledgments}
We use Gary Ferland for maintaining his freely distributed code {\tt CLOUDY}. 
We are grateful to A. Stamerra and J. Becerra for fruitful discussions.


\begin{thebibliography}{}

\bibitem[\protect\citeauthoryear{Abdo et al.}{2010}]{2010Natur.463..919A} Abdo A.~A., et al., 2010, Nature, 463, 919

\bibitem[\protect\citeauthoryear{Ackermann et al.}{2010}]{2010ApJ...721.1383A} Ackermann M., et al., 2010, ApJ, 721, 1383 

\bibitem[]{} Albert J., et al., 2008, Science, 320, 1752

\bibitem[\protect\citeauthoryear{Aleksi{\'c} et al.}{2011}]{2011ApJ...730L...8A} Aleksi{\'c} J., et al., 2011a, ApJ, 730, L8

\bibitem[\protect\citeauthoryear{Aleksi{\'c} et al.}{2011}]{2011arXiv1101.4645A} Aleksi{\'c} J., et al., 2011b, A\&A, 530, 4

\bibitem[\protect\citeauthoryear{Bentz et al.}{2009}]{2009ApJ...705..199B} Bentz M.~C., et al., 2009, ApJ, 705, 199 

\bibitem[\protect\citeauthoryear{B{\l}a{\.z}ejowski et al.}{2000}]{2000ApJ...545..107B} B{\l}a{\.z}ejowski M., Sikora M., Moderski  R., Madejski G.~M., 2000, ApJ, 545, 107 

\bibitem[\protect\citeauthoryear{Bonnoli et al.}{2011}]{2011MNRAS.410..368B} Bonnoli G., Ghisellini G., Foschini L., 
Tavecchio F., Ghirlanda G., 2011, MNRAS, 410, 368 

\bibitem[\protect\citeauthoryear{Decarli et 
al.}{2008}]{2008MNRAS.386L..15D} Decarli R., Dotti M., Fontana M., Haardt 
F., 2008, MNRAS, 386, L15

\bibitem[\protect\citeauthoryear{Decarli, Dotti, \& Treves}{2011}]{2011MNRAS.tmp..416D} Decarli R., Dotti M., Treves A., 2011, MNRAS, 416 

\bibitem[\protect\citeauthoryear{Denney et al.}{2009}]{2009ApJ...704L..80D} Denney K.~D., et al., 2009, ApJ, 704, L80 

\bibitem[\protect\citeauthoryear{Dermer \& Schlickeiser}{1993}]{1993ApJ...416..458D} Dermer C.~D., Schlickeiser R., 1993, ApJ, 416, 458 

\bibitem[\protect\citeauthoryear{Dermer et al.}{2009}]{2009ApJ...692...32D} Dermer C.~D., Finke J.~D., Krug H., B{\"o}ttcher M., 2009, ApJ, 692, 32

\bibitem[\protect\citeauthoryear{Dermer, Murase, \& Takami}{2012}]{2012ApJ...755..147D} Dermer C.~D., Murase K., Takami H., 2012, ApJ, 755, 147

\bibitem[\protect\citeauthoryear{Donea \& Protheroe}{2003}]{2003APh....18..377D} Donea A.-C., Protheroe R.~J., 2003, Astroparticle Phys., 18, 377

\bibitem[\protect\citeauthoryear{Elvis}{2000}]{2000ApJ...545...63E} Elvis M., 2000, ApJ, 545, 63 

\bibitem[\protect\citeauthoryear{Emmering, Blandford, \& Shlosman}{1992}]{1992ApJ...385..460E} Emmering R.~T., Blandford R.~D., Shlosman I., 1992, ApJ, 385, 460 

\bibitem[\protect\citeauthoryear{Ferland etal.}{1998}]{1998PASP..110..761F} Ferland G.~J., Korista K.~T., Verner
D.~A., Ferguson J.~W., Kingdon J.~B., Verner E.~M., 1998, PASP, 110, 761

\bibitem[\protect\citeauthoryear{Foschini et al.}{2010}]{2010MNRAS.408..448F} Foschini L., Tagliaferri G., Ghisellini  G., Ghirlanda G., Tavecchio F., Bonnoli G., 2010, MNRAS, 408, 448 

\bibitem[\protect\citeauthoryear{Frank, King, \& Raine}{2002}]{2002apa..book.....F} Frank J., King A., Raine D.~J., 2002, Accretion Power in Astrophysics, Cambridge Univ. Press, Cambridge, UK

\bibitem[\protect\citeauthoryear{Ghisellini \& Tavecchio}{2009}]{2009MNRAS.397..985G} Ghisellini G., Tavecchio F., 2009, MNRAS, 397, 985 

\bibitem[]{} Heitler, W., 1960 The Quantum Theory of Radiation, Oxford University Press, Oxford.

\bibitem[\protect\citeauthoryear{Jarvis \& McLure}{2006}]{2006MNRAS.369..182J} Jarvis M.~J., McLure R.~J., 2006, MNRAS, 369, 182

\bibitem[\protect\citeauthoryear{Kaspi \& Netzer}{1999}]{1999ApJ...524...71K} Kaspi S., Netzer H., 1999, ApJ, 524, 71

\bibitem[\protect\citeauthoryear{Kaspi et al.}{2000}]{2000ApJ...533..631K} 
Kaspi S., Smith P.~S., Netzer H., Maoz D., Jannuzi B.~T., Giveon U., 2000, 
ApJ, 533, 631 

\bibitem[\protect\citeauthoryear{Kaspi et al.}{2007}]{2007ApJ...659..997K} 
Kaspi S., Brandt W.~N., Maoz D., Netzer H., Schneider D.~P., Shemmer O., 
2007, ApJ, 659, 997 

\bibitem[\protect\citeauthoryear{Korista \& Goad}{2000}]{2000ApJ...536..284K} Korista K.~T., Goad M.~R., 2000, ApJ, 
536, 284 

\bibitem[\protect\citeauthoryear{Korista \& Goad}{2001}]{2001ApJ...553..695K} Korista K.~T., Goad M.~R., 2001, ApJ, 553, 695

\bibitem[\protect\citeauthoryear{Krolik}{1999}]{1999agnc.book.....K} Krolik J.~H., 1999, Active Galactic Nuclei: from the central black hole to the galactic environment, Princeton Univ. Press, Princeton, NJ

\bibitem[\protect\citeauthoryear{Liu \& Bai}{2006}]{2006ApJ...653.1089L} Liu H.~T., Bai J.~M., 2006, ApJ, 653, 1089 

\bibitem[\protect\citeauthoryear{Maraschi, Ghisellini, \& Celotti}{1992}]{1992ApJ...397L...5M} Maraschi L., Ghisellini G., Celotti A., 1992, ApJ, 397, L5

\bibitem[\protect\citeauthoryear{Marscher et al.}{2008}]{2008Natur.452..966M} Marscher A.~P., et al., 2008, Nature, 452, 966

\bibitem[\protect\citeauthoryear{Marscher et al.}{2010}]{2010ApJ...710L.126M} Marscher A.~P., et al., 2010, ApJ, 710, L126

\bibitem[\protect\citeauthoryear{McLure \& Dunlop}{2001}]{2001MNRAS.327..199M} McLure R.~J., Dunlop J.~S., 2001, MNRAS, 327, 199 

\bibitem[\protect\citeauthoryear{Nalewajko et al.}{2012}]{2012arXiv1202.2123N} Nalewajko K., Begelman M.~C., Cerutti B., Uzdensky D.~A., Sikora M., 2012, MNRAS, in press (arXiv:1202.2123) 

\bibitem[\protect\citeauthoryear{Netzer}{2008}]{2008NewAR..52..257N} Netzer H., 2008, NewAR, 52, 257 

\bibitem[\protect\citeauthoryear{Osterbrock}{1988}]{1988PASP..100..412O} Osterbrock D.~E., 1988, PASP, 100, 412

\bibitem[\protect\citeauthoryear{Peterson \& Wandel}{2000}]{2000ApJ...540L..13P} Peterson B.~M., Wandel A., 2000, ApJ, 540, L13

\bibitem[\protect\citeauthoryear{Pian et al.}{1999}]{1999ApJ...521..112P} Pian E., et al., 1999, ApJ, 521, 112 

\bibitem[\protect\citeauthoryear{Poutanen \& Stern}{2010}]{2010ApJ...717L.118P} Poutanen J., Stern B., 2010, ApJ, 717, L118 (PS10)

\bibitem[]{} Shields, G. A. 1978, in Proc. of the  Pittsburgh Conf. on BL Lac objects, Ed. A.M. Wolfe, p. 275

\bibitem[\protect\citeauthoryear{Sikora, Moderski, 
\& Madejski}{2008}]{2008ApJ...675...71S} Sikora M., Moderski R., Madejski G.~M., 2008, ApJ, 675, 71 

\bibitem[\protect\citeauthoryear{Sikora, Begelman, 
\& Rees}{1994}]{1994ApJ...421..153S} Sikora M., Begelman M.~C., Rees M.~J., 1994, ApJ, 421, 153 

\bibitem[\protect\citeauthoryear{Sikora et al.}{2009}]{2009ApJ...704...38S} 
Sikora M., Stawarz {\L}., Moderski R., Nalewajko K., Madejski G.~M., 2009, 
ApJ, 704, 38 

\bibitem[\protect\citeauthoryear{Sitarek \& Bednarek}{2008}]{2008MNRAS.391..624S} Sitarek J., Bednarek W., 2008, MNRAS, 391, 624

\bibitem[\protect\citeauthoryear{Stern \& Poutanen}{2011}]{2011MNRAS.417L..11S} Stern B.~E., Poutanen J., 2011, MNRAS, 417, L11 

\bibitem[\protect\citeauthoryear{Tanaka et al.}{2011}]{2011arXiv1101.5339T} Tanaka Y.~T., et al., 2011, ApJ, 733, 19
\bibitem[\protect\citeauthoryear{Tavecchio 
\& Ghisellini}{2008}]{2008MNRAS.386..945T} Tavecchio F., Ghisellini G., 2008, MNRAS, 386, 945 

\bibitem[\protect\citeauthoryear{Tavecchio \& Mazin}{2009}]{2009MNRAS.392L..40T} Tavecchio F., Mazin D., 2009, MNRAS, 392, L40 (TS09)

\bibitem[\protect\citeauthoryear{Tavecchio et al.}{2010}]{2010MNRAS.405L..94T} Tavecchio F., Ghisellini G., Bonnoli G., 
Ghirlanda G., 2010, MNRAS, 405, L94

\bibitem[\protect\citeauthoryear{Tavecchio et al.}{2011}]{2011arXiv1104.0048T} Tavecchio F., Becerra-Gonzales J., 
Ghisellini G., Stamerra A., Bonnoli G., Foschini L., Maraschi L., 2011, A\&A, 354, 86

\bibitem[\protect\citeauthoryear{Tavecchio et al.}{2012}]{2011arXiv1104.0048T} Tavecchio F., Roncadelli M., Galanti G. \& Bonnoli G., 2012, Phys. Rev. D, submitted, (arXiv:1202.6529)

\bibitem[\protect\citeauthoryear{Trump et al.}{2011}]{2011arXiv1103.0276T} Trump J.~R., et al., 2011, ApJ, 733, 60 

\bibitem[\protect\citeauthoryear{}{}]{}Wagner, S., \& Behera, B. 2010, in 10th HEAD Meeting, Hawaii (BAAS, 42, 2, 07.05)

\bibitem[\protect\citeauthoryear{Wills \& Browne}{1986}]{1986ApJ...302...56W} Wills B.~J., Browne I.~W.~A., 1986, ApJ, 302, 56 

\end{thebibliography}
\end{document}